\documentclass[12pt]{article}
\usepackage{epsfig}
\newcommand{\be}{\begin{equation}}
\newcommand{\en}{\end{equation}}
\newcommand{\eqa}{\begin{eqnarray}}
\newcommand{\ena}{\end{eqnarray}}
\newcommand{\half}{\frac{1}{2}}
\newcommand{\ra}{\rightarrow}

\newcommand{\tr}{\mbox{Tr}}
\newcommand{\noi}{\noindent}
\newcommand{\eq}[1]{(\ref{#1})}

\date{}
\begin{document}
\title{P-vortices, nexuses  and effects of gauge copies  \vskip-30mm
\rightline{\small ITEP-TH-12/2000}
\vskip 26mm
}
\author{V.G.~Bornyakov$^{\rm a}$, D.A.~Komarov$^{\rm b}$,
M.I.~Polikarpov$^{\rm b}$\\
and A.I.~Veselov$^{\rm b}$ \\
\\
$^{\rm a}$ {\small\it Institute for High Energy Physics, Protvino 142284,
Russia}\\
$^{\rm b}$ {\small\it Institute of Theoretical and  Experimental
Physics,}\\
{\small\it B.Cheremushkinskaya 25, Moscow, 117259, Russia}\\ }

\maketitle

\begin{abstract}
We perform the careful study of the gauge copies problem for
the direct center projection in $SU(2)$ lattice gauge theory. Our results
indicate that this gauge is not appropriate for the investigation of the
center vortices. We also show that the point--like objects, nexuses, are
important for the confinement dynamics.  \end{abstract}

\section{Introduction}

The old idea about the role of the center vortices in confinement phenomena
\cite{center_v} has been revived recently with the use of lattice
regularization. Both gauge invariant \cite{tomb1} and gauge dependent
\cite{greens1} approaches were developed. The gauge dependent studies were
done in a particular gauge, named center gauge.  Such gauge leaves intact
center group local gauge invariance. It is believed that gauge dependent
P-vortices defined on the lattice plaquettes are able to locate thick gauge
invariant center vortices and thus provide the essential evidence for the
center vortex picture of confinement. So far 3 different center gauges have
been used in practical computations: the indirect center gauge
\cite{greens1}, the direct center gauge \cite{greens2} and the Laplacian
center gauge \cite{def}. It is known that the first two of these gauges
suffer from gauge copies problem. Many results supporting the above
mentioned role of P-vortices were obtained in the direct center gauge.
Recently the following feature of this gauge has been discovered
\cite{tomb2}: there are gauge copies which correspond to higher maxima of
the gauge fixing functional $F$ (see below for definition) than usually
obtained and at the same time these new gauge copies produce P-vortices
evidently with no center vortex finding ability since projected Wilson loops
have no area law. It has been argued in \cite{greens3} that one can still
use direct center gauge to locate  center vortices if one uses gauge fixing
algorithm avoiding ``bad'' copies of \cite{tomb2}. Below we subject this
statement to the careful check. Another goal of our paper is
to investigate properties of recently introduced new objects called nexuses
\cite{cornwall,volovik} or center monopoles \cite{cpvz}. One can define
nexus in $SU(N)$ gauge theory as 3D object formed by $N$ center vortices
meeting at the center, or nexus, with the zero (mod $N$) net flux. We use
P--vortices in the center projection to define nexuses in $SU(2)$ lattice
gauge theory.

\section{Direct center gauge}

Direct center gauge is defined by the maximization of the following
functional of the lattice gauge field $U_{n,\mu}$ \cite{greens2}:

\be F(U) =
\frac{1}{4 V} \sum_{n,\mu} \left( \half\tr U_{n,\mu} \right)^2 = \frac{1}{4
       V} \sum_{n,\mu} \frac{1}{4}\left( \tr_{adj} U_{n,\mu} +1 \right),
     \label{maxfunc}
\en
\noi with respect to local gauge transformations, and can be considered as
Landau gauge for adjoint representation; $V$ is the lattice volume.
Condition (\ref{maxfunc}) fixes the gauge up to $Z(2)$ gauge transformation.
Fixed configuration can be decomposed into $Z(2)$ and coset parts:
$U_{n,\mu} = Z_{n,\mu} V_{n,\mu}$, where $Z_{n,\mu} = \mbox{sign} \tr
U_{n,\mu}$.  Plaquettes constructed from $Z_{n,\mu}$ field have values $\pm
1$. Those of them taking values $-1$ compose the so called P-vortices.
P-vortices form closed surfaces in 4D space. Some evidence has been
collected, that P-vortices in the direct center gauge can serve to locate
gauge invariant center vortices.  It has been reported \cite{greens2} that
projected Wilson loops computed via linking number of the static quarks
trajectories and P-vortices have area law with the string tension
$\sigma_{Z(2)}$ very close to the string tension of the nonabelian theory
$\sigma_{SU(2)}$. This fact has been called center dominance. Another
important observation was that the density of P-vortices scales as a
physical quantity \cite{greens2,tubing}.  We inspect these statements using
careful gauge fixing procedure.

The most common method to fix the gauge of the type (\ref{maxfunc}) is the
relaxation algorithm which makes maximization iteratively site by site. The
relaxation is made more effective with the help of the overrelaxation. It is
known that another algorithm -- simulated annealing -- is more effective and
very useful when gauge copies problem becomes severe \cite{bbms}.  Here we
do not employ simulated annealing and apply gauge fixing procedure explained
in details in ref. \cite{greens2}. We call it RO (relaxation --
overrelaxation) procedure.

The main problem of the direct center gauge fixing is that the functional
$F(U)$ \eq{maxfunc} has many local maxima. We call configurations
corresponding to these local maxima gauge copies. They are lattice Gribov
copies in fact. It is well known that for some gauge conditions which are
formulated as the maximization of a nonlocal functional (e.g. Landau,
Coulomb and Maximal Abelain gauges) the gauge dependent quantities depend
strongly on the local maxima picked up, while to find out the global maximum
is impossible. Thus it is necessary to approach the global maximum as close
as possible. We follow the following procedure proposed and checked in
\cite{bbms}: for given configuration we generate $N_{cop}$ gauge equivalent
copies applying random gauge transformations, and fix the gauge for each
gauge copy using the RO procedure.  After that we compute the gauge
dependent quantity $X$ on the gauge copy corresponding to the highest
maximum of \eq{maxfunc}, $F_{max}(N_{cop})$. Averaging over
statistically independent gauge field configurations and varying $N_{cop}$
we obtain the function $X(N_{cop})$ and extrapolate it to $N_{cop} \ra
\infty$ limit. This should provide a good estimation for $X$ computed on
the global maximum unless the algorithm in use does not permit to reach the
global maximum or its vicinity (the situation we met also in the present
study). The main difference of the present study from the calculations
performed earlier is that we use the higher value of the gauge copies
($1 \leq N_{cop} \leq 20$) than it was used in refs.
\cite{greens1,greens2,greens3,tubing} and make careful analysis of $N_{cop}$
dependence. Due to that our results differ drastically from those reported
previously \cite{greens1,greens2,greens3,tubing}.

Separately we compute observables using the modified (LRO) gauge fixing
procedure \cite{tomb2}: every configuration has been first fixed to Landau
gauge, and then the RO algorithm for the direct center gauge has been
applied. In this case the effect of large number of gauge copies, $N_{cop}$,
is not very important, we confirm the results of ref. \cite{tomb2}.

Note that there exists another proposal \cite{parfac} for the general gauge
fixing procedure which is free of gauge copies problem. In some particular
limit this procedure corresponds to the search of the global maximum
\cite{bbms}.  There is also a class of gauge conditions \cite{def},
\cite{vds} which do not suffer from the gauge copies problem.

\section{Results}

Our computations have been performed on lattice $L^4=12^4$ for $\beta=2.3,
2.4$ and $L^4=16^4$ for $\beta=2.5$. For $\beta = 2.3, 2.4$ ($\beta = 2.5$)
we study $100$ ($50$) statistically independent gauge field configurations.
Using the described above gauge fixing procedure we calculate the various
observables as functions of the number of randomly generated gauge copies
$N_{cop}$ ( $1 \leq N_{cop} \leq 20$).

{\it (i)} We confirm the conclusion of ref. \cite{tomb2} that gauge copies
generated via LRO procedure have higher maxima of $F(U)$ and thus are closer
to the global maximum of $F(U)$. We found that $F_{max}^{LRO}(N_{cop})
> F_{max}^{RO}(N_{cop})$ for any value of $N_{cop}$, at any considered value
of $\beta$.

{\it (ii)} We find that LRO procedure gives copies
with significantly lower density, $\rho$, of P-vortices than RO
procedure. We use the standard definition: $\rho = \frac{1}{12 \cdot V}
\sum_{n;\mu > \nu} (1-Z_{n,\mu\nu})$. Thus gauge copies generated by RO and
LRO procedures are indeed different even in the limit $N_{cop} \ra \infty$.

{\it (iii)} The difference between LRO and RO procedure results can be
qualitatively explained as follows. Fixing the Landau gauge we get the
configuration almost without P-vortices, the subsequent RO procedure
substantially increases the number of P-vortices but percolating cluster
does not appear. The original gauge field configuration contains a lot of
P-vortices and the local RO procedure is not able to remove all large (and
even wrapping) clusters of P-vortices. The field configuration after
application of LRO procedure contains many small P-vortex clusters; the
field configuration after application of RO procedure contains one large
percolating cluster. It seems that this cluster is responsible for the area
law behavior of the projected Wilson loops (see below).

{\it (iv)} The most important observable is the $Z(2)$-projected Creutz ratio
$\chi (I)$ which we calculate using the procedure suggested in refs.
\cite{greens1}, \cite{greens2}. $\chi (I)$ is defined through the projected
Wilson loops, $W_{Z(2)}(C)=\exp\{i\pi {\cal L}(\Sigma_P,C)\}$. Here ${\cal
L}(\Sigma_P,C)$ is the 4D linking number of the closed surface, $\Sigma_P$,
formed by P-vortex and closed loop $C$.

\begin{figure}[tbh]
\begin{centering}
\epsfig{figure=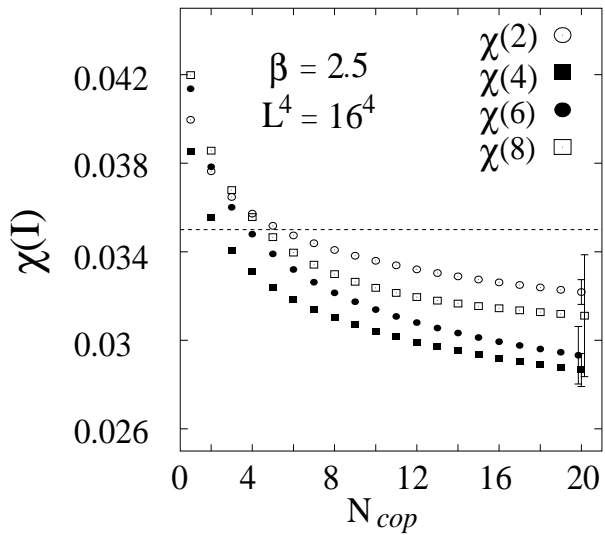,height=7cm,width=8cm}
\caption{ The dependence of the Creutz ratios $\chi (I)$
on the number of gauge
copies $N_{cop}$ for $\beta=2.5, L^4=16^4$.  The error bars are shown for one
value of $N_{cop}$ only and they are characteristic for the other data
points. The dashed line corresponds to the nonabelian string tension,
$\sigma_{SU(2)}$.}
\end{centering}
\end{figure}
In Fig.1. we show the dependence of $\chi (I)$ on $N_{cop}$ for
$\beta=2.5$. It occurs that this dependence is nicely fitted by the function
$C_1 + C_2/\sqrt{N_{cop}}$. The reason for such dependence is still to be
understood.  In Table 1 we give the ratio $\sigma_{Z(2)}/\sigma_{SU(2)}$
\footnote{The data for $\sigma_{SU(2)}$ are taken from \cite{bss}}.
$\sigma_{Z(2)}$ is computed from $\chi (I)$ for $3 \leq I \leq 4$
data at $12^4$ lattice and for $3 \leq I \leq 6$ data at $16^4$ lattice.
For $N_{cop}=3$ (number of gauge copies used in \cite{greens2})
$\sigma_{Z(2)}$ is close to $\sigma_{SU(2)}$. But it becomes significantly
lower for $N_{cop} \rightarrow \infty$. Thus RO procedure results strongly
depend on $N_{cop}$. It is important that $\sigma_{Z(2)}$ is 20-30 \% lower
than $\sigma_{SU(2)}$ for $N_{cop} \to \infty$. This implies that even if
one restricts oneself to RO procedure as it is suggested in \cite{greens3},
one cannot conclude that P-vortices indeed well locate {\it all} center
vortices.

{\it (v)} For gauge copies generated by LRO procedure we confirm the result
of \cite{tomb2} that $\chi (I)$ is zero within statistical errors for any
value of $N_{cop}$.

{\it (vi)} In Table 1 we also show the ratio $2\rho/\sigma_{SU(2)} a^2$
($\rho$ is the density of P-vortices). As it is claimed in ref.
\cite{tubing} in case of the uncorrelated plaquettes
carrying P-vortices $2 \rho$ coincides with the dimensionless string
tension, $\sigma_{SU(2)} a^2$. The results presented in Table~1 show that
the density of P-vortices is not proportional to $\sigma_{SU(2)} a^2$.
We have found out that for $N_{cop} = 3 \quad 
\rho$ is in good agreement with the asymptotic scaling as it was found in
\cite{greens2}. But for  $N_{cop} \to \infty \quad \rho$ deviates from the
two loop asymptotic scaling formula.

\begin{table}[tp]
\caption{The comparison of $\sigma_{Z(2)}$, $\sigma_{SU(2)}$ and $\rho$ for
RO gauge fixing center projection. } \label{t1} \vspace{0.5cm}
\setlength{\tabcolsep}{0.55pc}
\begin{centering}
\begin{tabular}{c|ccc|ccc}  \hline
$N_{cop}$& & $ \sigma_{Z(2)}/\sigma_{SU(2)}$ & && $2\rho/(\sigma_{SU(2)}
a^2)$ & \\ \hline &$\beta=2.3$ &$\beta= 2.4$ &$ \beta=
2.5$&$\beta=2.3$&$\beta=2.4$&$\beta=2.5$   \\ \hline 3&0.94(2) &0.93(2) &
0.98(2)     &1.30(1) &1.51(1) & 1.74(1) \\ 20&0.87(2)  &0.80(2)  & 0.83(3)
&1.27(1) &1.42(1) & 1.61(2)\\ $\infty$&0.82(3)&0.71(3)& 0.71(3)&1.24(1)
&1.33(2) & 1.49(2) \\ \hline \end{tabular} \end{centering} \end{table}

{\it (vii)} We also investigate the properties of the point like objects,
called nexuses. On the 4D lattice we have the conserved currents of nexuses,
defined after the center projection.  We calculate the phase, $s_l$, of the
$Z(2)$ link variable:  $Z_l = \exp(i\pi s_l), ~s_l=0,1$.  Then we define the
plaquette variable $\sigma_{P} = \mbox{d}s~~ \mbox{mod}~ 2$, $(\sigma_P =
0,1)$. The nexus current (or center monopole current \cite{cpvz}) is then
defined as $^{\star}j=\frac{1}{2}\delta^{\star}\sigma_P$. These currents
live on the surface of the P-vortex (on the dual 4D lattice) and P-vortex
flux goes through positive and negative nexuses in alternate order. The
important characteristic of the cluster of currents is the condensate, $C$,
defined \cite{polikarpov} as the percolation probability. As it is shown in
ref. \cite{cpvz} the condensate $C$ of the nexus currents is the order
parameter for the confinement -- deconfinement phase transition. We
found that $C$ is nonzero for the gauge copies obtained via RO procedure
(when the projected Wilson loops have the area law). $C$ is zero (in the
thermodynamic limit $L \rightarrow \infty$) for gauge copies obtained using
LRO procedure (when the projected Wilson loops have no area law). It is
interesting that for RO procedure $C$ seems to scale as the physical
quantity with the dimension $(mass)^4$. This is illustrated in Fig.2, where
we plot the $\beta$--dependence of the ratio $C/(\sigma_{SU(2)} a^2)^2$.
Thus these new objects might be important degrees of freedom for the
description of the nonperturbative effects.

\begin{figure}
\begin{centering}
\epsfig{figure=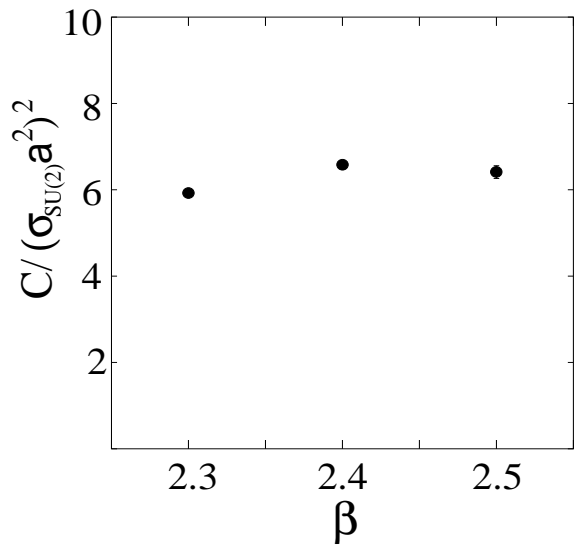,height=7cm,width=7.5cm}
\caption{The $\beta$--dependence of the ratio of the nexus condensate, $C$,
to the $SU(2)$ string tension in lattice units \cite{bss}.}
\end{centering}
\end{figure}

{\it (viii)} It is important to perform the same calculations for
the indirect center gauge \cite{greens1} and for the Laplacian center gauge
\cite{def}.

We thank Ph. de Forcrand and T. Kovacs for useful remarks.
This study was partially supported by grants RFBR 96-15-96740, RFBR
99-01230a, INTAS 96-370 and Monbushu grant.

\end{document}